\documentclass[prd,preprintnumbers,superscriptaddress,nofootinbib,twocolumn]{revtex4-2}
\usepackage{epsfig}
\usepackage{amsmath}
\usepackage{amsfonts}
\usepackage{float}
\usepackage{amssymb}
\usepackage{slashed}
\usepackage[dvipsnames]{xcolor}
\usepackage{pbox}
\usepackage{subfigure}
\usepackage[colorlinks,citecolor=blue, linkcolor = blue, urlcolor = blue]{hyperref}
\usepackage{tabularx}
\usepackage{titlesec}
\usepackage{scrextend}
\usepackage[normalem]{ulem}
\usepackage{orcidlink}
\usepackage{natbib}

\def\bal#1\eal{\begin{align}#1\end{align}}
\def\beq#1\eeq{\begin{equation}#1\end{equation}}

\def\pim{{\pi^-}}
\def\pip{{\pi^+}}
\def\MeV{{\rm MeV}}

\def\gap{g_{ap}}
\def\gan{g_{an}}
\def\gagg{g_{a \gamma \gamma}}

\def\del{\partial}
\def\nn{\nonumber}

\def\Lag{\mathcal{L}}
\def\Deltaplus{\Delta^{+}_{\mu}}
\def\Deltazero{\Delta^{0}_{\mu}}
\def\Deltabarplus{\bar{\Delta}^{+}_{\mu}}
\def\Deltabarzero{\bar{\Delta}^{0}_{\mu}}

\begin{document}
\title{Fresh look at the diffuse ALP background from supernovae}

\author{Francisco R. Candón\,\orcidlink{0009-0002-3199-9278}}
\email{francandon@unizar.es}
\affiliation{Centro de Astropartículas y Física de Altas Energías, University of Zaragoza, Zaragoza, 50009, Aragón, Spain}

\author{Sougata Ganguly\,\orcidlink{0000-0002-8742-0870}}
\email{sganguly0205@ibs.re.kr}
\affiliation{Particle Theory  and Cosmology Group, Center 
for Theoretical Physics of the Universe,\\
Institute for Basic Science (IBS), Daejeon, 34126, Korea}

\author{Maurizio  Giannotti\,\orcidlink{0000-0001-9823-6262}}
\email{mgiannotti@unizar.es}
\affiliation{Centro de Astropartículas y Física de Altas Energías, University of Zaragoza, Zaragoza, 50009, Aragón, Spain}
\affiliation{Physical Sciences, Barry University, 11300 NE 2nd Avenue, Miami Shores, Florida 33161, USA}

\author{Tanmoy Kumar\,\orcidlink{0000-0001-9775-6645}}
\email{kumartanmoy1998@gmail.com}
\affiliation{School of Physical Sciences, Indian Association for the Cultivation of Science, 2A \& 2B Raja S.C. Mullick Road, Jadavpur, Kolkata 700032, India}

\author{Alessandro Lella\,\orcidlink{0000-0002-3266-3154}}
\email{alessandro.lella@ba.infn.it}
\affiliation{Dipartimento Interateneo di Fisica  ``Michelangelo Merlin'', Via Amendola 173, 70126 Bari, Italy}
\affiliation{Istituto Nazionale di Fisica Nucleare - Sezione di Bari, Via Orabona 4, 70126 Bari, Italy}%

\author{Federico Mescia\,\orcidlink{0000-0003-3582-2162}}
\email{federico.mescia@lnf.infn.it}
\affiliation{Istituto Nazionale di Fisica Nucleare, Laboratori Nazionali di Frascati, \\ C.P.~13, 00044 Frascati, Italy}

\begin{abstract}
Protoneutron stars, highly compact objects formed in the core of exploding supernovae (SNe), are powerful sources of axionlike particles (ALPs). In the SN core, ALPs are dominantly produced via nucleon-nucleon bremsstrahlung and pion conversion, resulting in an energetic ALP spectrum peaked at energies 
${\cal O} (100) \,\rm MeV$. In this work, we revisit the diffuse ALP background, produced from all past core-collapse supernovae, and update the constraints derived from {\it Fermi}-LAT observations. Assuming the maximum ALP-nucleon coupling allowed by the SN 1987A cooling, we set the upper limit $g_{a \gamma \gamma} \lesssim 2 \times 10^{-13}\,\rm GeV^{-1}$ for ALP mass $m_a\lesssim 10^{-10}\,\rm eV$,
which is approximately a factor of two improvement with respect to the existing bounds.
On the other hand, for $m_a \gtrsim 10^{-10}\,\rm eV$, we find that including pion conversion strengthens the bound on $g_{a\gamma \gamma}$, approximately by a factor of two compared to the constraint obtained from bremsstrahlung alone.
Additionally, we present a sensitivity study for future experiments such as AMEGO-X, e-ASTROGAM,  GRAMS-balloon, GRAMS-satellite, and MAST. We find that the expected constraint from MAST would be comparable to {\it Fermi}-LAT bound. However, SN 1987A constraint remains one order of magnitude stronger as compared to the bound derived from the current and future gamma-ray telescopes.

\end{abstract}
\preprint{BARI-TH/777-25}
\preprint{CTPU-PTC-25-12}
\maketitle
\section{INTRODUCTION}
\label{sec:introduction}
The core collapse of massive stars leads to the formation of a protoneutron star (PNS), a hot and dense astrophysical environment capable of producing axionlike particles (ALPs). Core-collapse supernovae (SNe) may act as cosmic factories of axions and axionlike particles  (ALPs)~\cite{Raffelt:1996wa,Raffelt:2006cw,Fischer:2016cyd, Caputo:2024oqc}, providing an opportunity to probe these particles and their couplings using a variety of techniques.

These weakly interacting pseudoscalar particles, if sufficiently light, can be abundantly produced and escape the PNS, contributing to a diffuse background of SN ALPs, analogous to the diffuse supernova neutrino background. Historically, supernovae have played a crucial role in constraining ALP properties, particularly following the detection of the SN 1987A neutrino burst. Indeed, the SN 1987A neutrino detection was a milestone event for axion physics  
(see Refs.~\cite{Carenza:2024ehj,Carenza:2023lci} for recent reviews).

\begin{figure}[h!]
    \centering
    \includegraphics[width=0.48\textwidth]{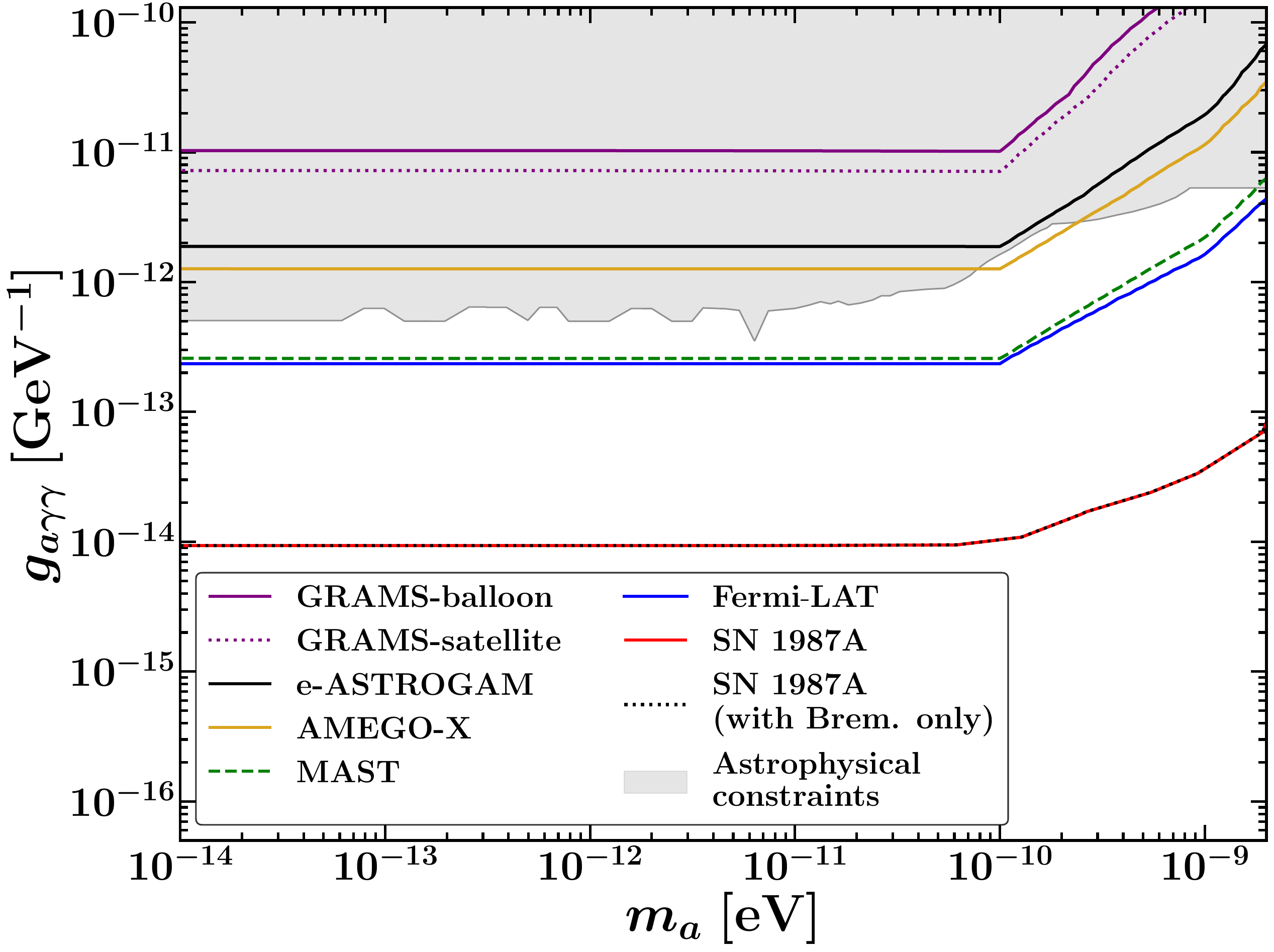}
    \caption{Projected constraints in the $m_a - g_{a \gamma \gamma}$ plane from future gamma ray missions AMEGO-X \cite{Caputo:2022xpx}, e-ASTROGAM \cite{e-ASTROGAM:2016bph}, GRAMS-balloon \cite{Aramaki:2019bpi}, GRAMS-satellite \cite{Aramaki:2019bpi} and MAST \cite{Dzhatdoev:2019kay}.
    Here we have considered supernova ALP production via both bremsstrahlung and pion conversion. 
    For deriving the above constraints, we have assumed $g_{ap} = 5 \times 10^{-10}$ and $g_{an} = 0$. The constraint from {\it Fermi}-LAT is shown by the blue solid line. The solid red line denotes the SN 1987A constraint considering both bremsstrahlung and pion conversion, whereas the case of ALP production only via bremsstrahlung is shown with the black dotted line.
    The gray shaded region represents combined astrophysical constraint on $g_{a \gamma \gamma}$ considering ALPs not to be a dark matter candidate \cite{AxionLimits}.
    }
    \label{fig:future_constraints}
\end{figure}

The axion emission leads to an additional energy loss, which could, in principle, have an observable impact on the detected neutrino signal. This argument led to very stringent limits on the ALP-nucleon interactions.
Primarily, the focus has been on the nucleon-nucleon ($NN$) bremsstrahlung process~\cite{Turner:1987by,Burrows:1988ah,Burrows:1990pk,Raffelt:1987yt,Raffelt:1990yz,Keil:1996ju,Hanhart:2000ae,Giannotti:2005tn, Carenza:2019pxu, Choi:2021ign} as the dominant production mechanism. 
However, more recently, the ALP production from negatively charged pions via $\pi^-+ p \to a+n$, has emerged as the potentially dominant process~\cite{Carenza:2020cis,Fischer:2021jfm,Lella:2022uwi, Ho:2022oaw,Lella:2023bfb}.


If ALPs interact exclusively with photons, their production in supernovae occurs through the Primakoff process. However, this mechanism is relatively inefficient, making the resulting cooling bound less competitive compared to constraints from globular clusters~\cite{Ayala:2014pea, Straniero:2015nvc, Dolan:2022kul} (which are applicable for ALP masses $\lesssim 50$ keV~\cite{Cadamuro:2011fd,Carenza:2020zil, Lucente:2022wai}). Nonetheless, light ALPs produced in SNe can convert into gamma rays while traversing through the magnetic field of the Milky Way~\cite{Grifols:1996id,Brockway:1996yr}. 
The absence of a gamma-ray signal coincident with the SN 1987A neutrino burst in the data from the Gamma-Ray Spectrometer (GRS) aboard the Solar Maximum Mission (SMM) led to stringent constraints on the ALP-photon coupling~\cite{Grifols:1996id,Brockway:1996yr}. 
In particular, the most recent analysis performed for ALPs coupling only to photons reports an upper bound on $g_{a\gamma\gamma} < 5.3 \times 10^{-12}$ GeV$^{-1}$ for $m_a < 4 \times 10^{-10}$~eV~\cite{Payez:2014xsa} (see also the revision in~\cite{Hoof:2022xbe}).

Future Galactic supernovae occurring during the operational lifetime of the Large Area Telescope aboard the {\it Fermi}~satellite ({\it Fermi}-LAT) could significantly enhance these constraints~\cite{Meyer:2020vzy, Calore:2021hhn, Calore:2023srn}. The sensitivity to light ALPs converting into gamma rays in the Galactic magnetic field would be further improved if ALPs also couple to nucleons, as nucleon-induced processes can boost the ALP production rate~\cite{Lella:2024hfk}. Additionally, the detection of a gamma-ray signal in {\it Fermi}-LAT would provide valuable insights into the SN explosion mechanism, allowing one to probe the properties of the SN core and assess the relevance of the pion-induced production channel for ALPs. Ultimately, this could shed light on the equation of state of dense nuclear matter~\cite{Lella:2024hfk}.\footnote{The direct detection of SN ALPs from Earth, which do not rely on the conversion in the galactic magnetic field, has also been explored in a series of recent studies~\cite{Carenza:2023wsm,Asai:2022pio,Arias-Aragon:2024gdz,Ge:2020zww,Li:2023thv,Carenza:2025uib,Alonso-Gonzalez:2024ems}.}
 
Finally, Ref.~\cite{Calore:2020tjw} adopted a significantly different approach and investigated the diffuse ALP flux originating from past core-collapse SNe. 
The study showed that the cumulative emission of ALPs from all past core-collapse events gives rise to a diffuse flux with characteristic energies of ${{\mathcal O}(50)~{\rm MeV}}$. The nondetection of a diffuse supernova ALP background (DSALPB) was used to constrain ALP couplings to photons and nucleons, leveraging measurements of the diffuse gamma-ray flux obtained by the {\it Fermi}-LAT telescope. 
As expected, the inclusion of ALP-nucleon interactions significantly enhances the ALP production rate in SNe via the nucleon-nucleon bremsstrahlung process. 
Assuming the largest phenomenologically allowed ALP-nucleon coupling, the upper bound $g_{a\gamma\gamma} \lesssim 6 \times 10^{-13}$~GeV$^{-1}$ for $m_a \lesssim 10^{-11}$ eV was derived.\footnote{Constraints on 
$\gagg$  at higher ALP masses, incorporating the pionic contribution, from the nonobservation of the DSALPB by \textit{Fermi}-LAT have been discussed in Ref.~\cite{Benabou:2024jlj}.}

In this work, we revisit the analysis presented in Ref.~\cite{Calore:2020tjw} by incorporating key improvements that refine the constraints on the DSALPB. First, we account for the additional ALP production via the pion-induced process, which is now recognized as the dominant production mechanism in supernovae. 
This represents a significant advancement over previous studies, as it modifies both the expected ALP flux and its spectral distribution. As expected, the inclusion of the pion-induced production channel significantly alters the expected DSALPB spectrum, leading to a spectral hardening, as illustrated in Fig.~\ref{fig:dsalpb_flux}. 
In particular, the peak energy shifts to $\gtrsim 100$~MeV, modifying both theoretical expectations and the experimental requirements for detection.

Second, we assess the sensitivity of future (proposed) observational campaigns to detect the DSALPB signal. 
Our results are summarized in Fig.\,\ref{fig:future_constraints}. 
As we shall show, and as evident from Fig.~\ref{fig:future_constraints}, {\it only} MAST would have the required sensitivity to be competitive with {\it Fermi}-LAT.
Hence, it is unlikely that any of the future telescopes, in their currently proposed configurations, would be able to make a breakthrough in DSALPB searches.
We believe that this is an additional motivation for the next-generation MeV-range missions, which could search for the DSALPB and explore previously inaccessible regions of ALP parameter space and significantly enhance the prospects for discovery.

The paper is organized as follows:
In Sec.~\ref{sec:lagrangian}, we introduce the effective Lagrangian governing ALP interactions.
In Sec.~\ref{sec:ALP_production}, we describe the production mechanisms in supernovae for ALPs coupled to nuclear matter
and compute the resulting DSALPB flux.
Section~\ref{sec:gamma_ray_dsalpb} discusses ALP-photon conversion in the Milky Way’s magnetic field
and the expected gamma-ray signal.
Finally, in Sec.~\ref{sec:constraints}, we present our results on the projected constraints
on ALP couplings from future observations and conclude in Sec.~\ref{sec:conclusions}.

\section{EFFECTIVE LAGRANGIAN OF ALP}
\label{sec:lagrangian}
The phenomenological ALP Lagrangian relevant for our study can be written as 
\beq
\Lag_a = \dfrac{1}{2}\,\partial_\mu\,a\,\partial^\mu\,a - \dfrac{1}{2}\,m_a^2\,a^2 - 
\dfrac{g_{a \gamma \gamma}}{4} a F_{\mu \nu} \tilde{F}^{\mu \nu}
+\Lag_{\rm nuc},
\label{eq:alp_lagrangian}
\eeq
where $m_a$ is the mass of the ALP and  $g_{a\gamma \gamma}$ is the coupling of the ALP with the photon. 
The last term $\Lag_{\rm nuc}$ contains the interaction of ALP with hadrons as well as other hadronic interaction vertices relevant for ALP production processes.
Additional possible interaction terms, for example with electrons and positrons, will be ignored in this study. 
The explicit form of $\Lag_{\rm nuc}$ is given by \cite{Choi:2021ign, DiLuzio:2020wdo, Lella:2022uwi} 
\bal
\label{eq:alp_nucleon_lagrangian}
\Lag_{\rm nuc}  =\, &\dfrac{\del^\mu a}{2 m_N} 
\left[
g_{ap}\, \bar{p} \gamma_\mu \gamma_5 p + g_{an}\, \bar{n} \gamma_\mu \gamma_5 n \right.\nn\\
&\left.+\dfrac{g_{a \pi N}}{f_\pi} \left(\, i \pi^+ \bar{p} \gamma_\mu n-i \pi^- \bar{n} \gamma_\mu p \right) \right.\nn\\
&\left.+ g_{a N \Delta} \left(\, \bar{p} \Deltaplus + \Deltabarplus p\, 
+ \bar{n}\Deltazero + \Deltabarzero n \right) 
\right] \nn\\
&+ \dfrac{g_A}{2f_\pi}
\left[
\del_\mu \pi^0 \left(\bar{p} \gamma^\mu \gamma_5 p - \bar{n} \gamma^\mu \gamma_5 n\right) \right.\nn\\
 &\left.+\sqrt{2} \del_\mu \pip \bar{p} \gamma^\mu \gamma_5 n
+ \sqrt{2} \del_\mu \pim \bar{n} \gamma^\mu \gamma_5 p
\right]\,\,,
\eal
where nucleon mass $m_N = 0.938\,\rm GeV$, $g_A = 1.28$, and the pion decay constant $f_\pi = 92.4\,\rm MeV$. The first two terms
are the trilinear interactions of ALP with proton and neutron with coupling coefficients $\gap$ and $\gan$, respectively.
The four-particle interaction terms among ALP, pion, neutron, and proton are given in the second line of Eq.\,\eqref{eq:alp_nucleon_lagrangian} with coupling coefficient $g_{a \pi N} = (\gap - \gan)/\sqrt{2} g_A$ \cite{Choi:2021ign}
whereas ALP-$\Delta$ baryons interactions are written in the third line and $g_{a N \Delta} = -\sqrt{3}/2 (\gap - \gan)$ \cite{Ho:2022oaw}.
The relevant interactions between pion and nucleons in the context of SN physics are described in the last two lines
of Eq.\,\eqref{eq:alp_nucleon_lagrangian}.


In Eq.\,\eqref{eq:alp_nucleon_lagrangian}, 
the interaction terms between the ALP and neutrons, protons, and negatively charged pions facilitate ALP production from the nuclear plasma within the PNS,
as discussed in Ref.\,\cite{Lella:2023bfb}. In the next section, we briefly review the production of ALP from a PNS.

\section{ALP PRODUCTION FROM PNS}
\label{sec:ALP_production}
In the presence of ALP-nucleon interactions, ALPs can be copiously produced from the hot and dense environment of a PNS, primarily via \textit{NN}-bremsstrahlung and pion conversion.\footnote{In the SN core, ALPs can also be produced via Primakoff process \cite{Calore:2020tjw}. However, the ALP flux from the Primakoff process is noncompetitive with respect to the bremsstrahlung and pion conversion processes for the values of $g_{a\gamma\gamma}$ allowed by other bounds. } 

The rate of ALP production via \textit{NN}-bremsstrahlung per unit volume and per unit ALP energy interval can be calculated as \cite{Carenza:2023lci}
\bal
\left(\dfrac{d^2 n_a}{ d E_a dt} \right)_{NN}
 &= \dfrac{p_a}{4 \pi^2} 
 \int \left[\prod_{i=1}^4 d \Pi_i\right]\nn\\
 & \times (2 \pi)^4 \delta^4 \left(P_1 + P_2 - P_3 - P_4 - P_a\right) \nn\\
 & \times S|\mathcal{M}|_{NN}^2 \, 
 f_1 f_2 (1 - f_3) (1-f_4)\,\,\,,
 \label{eq:brem_flux_general}
\eal
where the four momenta of initial and final state nucleons are $P_i = (E_i, \vec{p}_i)$ with $(i =1,..,4)$.
$P_a = (E_a, \vec{p}_a)$ is the four momentum of ALP and $p_a \equiv |\vec{p}_a|$. $d\Pi_i = {d^3 \vec{p}_i}/{(2 \pi)^3 2 E_i}$ is the Lorentz invariant phase space measure, $|\mathcal{M}|_{NN}^2$ is the matrix amplitude square of the bremsstrahlung process, summed over initial and final spins
, and $S$ is the symmetry factor for the identical particles in the initial and final states. $f_i$ $(i =1,..,4)$ are the energy distribution
functions of the initial and final state nucleons and their explicit form is 
\bal
\label{eq:dist_fn}
f_i (E_i) = \left[\exp\left(\dfrac{E_i - \mu_i}{T_{SN}}\right) + 1\right]^{-1}\,\,,
\eal
where $T_{SN}$ is the temperature of the PNS and $\mu_i$ is the chemical potential.
Inside the PNS, $T_{\rm SN} \sim 30\,\MeV$ and thus protons and neutrons are nonrelativistic. Assuming negligible recoil momenta of the nucleons, the magnitude of their 3-momentum is $\sim \sqrt{m_N T_{SN}} \approx 168\,\rm MeV$, whereas 
for $T_{SN}\gg m_a$, $p_a \sim T_{SN}$. Since $p_a \ll |\vec{p}_i|$, we neglect $\vec{p}_a$ with respect to $\vec{p}_i$ in the momentum
conservation condition in Eq.\,\eqref{eq:brem_flux_general}~\cite{Brinkmann:1988vi,Giannotti:2005tn,Choi:2021ign}.

ALPs can also be produced from negatively charged pions via $\pi^- + p \to n + a$ process. We can write the ALP production rate per unit volume and per unit ALP energy interval as \cite{Carenza:2023lci}
\bal
\label{eq:pion_flux_general}
\left(\dfrac{d^2 n_a}{ d E_a dt} \right)_{\pi N}
 &= \dfrac{p_a}{4 \pi^2} 
 \int d \Pi_{\pi} d \Pi_n d \Pi_p \nn \\
 & \times (2 \pi)^4  \times\delta^4 \left(P_p + P_\pi - P_n - P_a \right) \nn\\
& \times |\mathcal{M}|^2_{\pi N} f_\pi f_p (1 - f_n)\,\,,
\eal
where $P_n, P_p$, and $P_\pi$ are the four momentum of neutron, proton, and pion respectively. The energy distribution
functions of protons and neutrons are denoted by $f_p$ and $f_n$, respectively, and their analytical form is given in Eq.\,\eqref{eq:dist_fn}.
The energy distribution function of negatively charged pions is
\bal
f_\pi =\left[\exp\left(\dfrac{E_\pi - \mu_\pi}{T_{SN}}\right) - 1\right]^{-1},
\eal
where $\mu_\pi$ is the chemical potential of the negatively charged pions inside the PNS and it is calculated using the $\beta$-equilibrium condition, $\mu_\pi = \mu_n - \mu_p$~\cite{Fore:2019wib}.
Finally $|\mathcal{M}|^2_{\pi  N}$ is the matrix amplitude square of $\pi^- + p \to n + a$,
summed over initial and final state spins. Since nucleons are nonrelativistic in SN environment, we approximated ${E_n \sim E_p}$
in the calculation of ALP production rate from pion conversion, so that  $E_a \simeq E_{\pi}$ where 
$E_\pi$ is the energy
of negatively charged pion.
Hence the total production rate of ALPs inside a newly born PNS per unit volume and per unit ALP energy interval is the sum of the above two contributions
\bal
\left(\dfrac{d^2 n_a}{ d E_a dt} \right)_{\rm total} = 
\left(\dfrac{d^2 n_a}{ d E_a dt} \right)_{NN} + \left(\dfrac{d^2 n_a}{ d E_a dt} \right)_{\pi N}\,\,.
\eal

Once produced, ALPs with sufficiently high kinetic energies can escape the PNS provided that the ALP-nucleon coupling is not very large such that the scattering rate of ALPs with the nucleons in the PNS medium is small. This regime is known as \emph{free-streaming regime}. 
Ref.~\cite{Lella:2023bfb} pointed out that ALPs with nuclear couplings $g_{aN}\lesssim10^{-8}$ would stream out from the SN core without relevant absorption effects.
While traveling out of the PNS, the produced ALPs have to overcome the strong gravitational potential generated in the densest regions of the PNS. As a result, their energy spectrum gets redshifted, i.e., an ALP that is produced inside the PNS with energy $E_a$, after escaping the PNS, is left with an energy $E^*_a$ where
\beq
E_a^* = \alpha(r) E_a\,\,,
\eeq
with $\alpha(r)$ being the lapse factor that encodes the redshifting of the energy in the gravitational potential of the PNS. The values
of $\alpha$ and other SN parameters such as density, temperature, and chemical potential at each radius are obtained from the SN simulation model SFHo-s18.8 with progenitor mass $18.8\,M_\odot$
by the \texttt{GARCHING} group \cite{SNarchive}, which was also previously used in~\cite{Lella:2022uwi,Lella:2023bfb,Lella:2024dmx,Lella:2024hfk,Carenza:2025uib}. As discussed in \cite{Carenza:2025uib}, in this model, the highest temperature ($\sim 40\,\rm MeV$) is achieved during the early cooling phase i.e., between $t = 1-2$ s after which it decreases. The core density, on the other hand, increases from $6 \times 10^{14}\,\rm gm/cm^3$ at $t = 1$ s to  $8 \times 10^{14}\,\rm gm/cm^3$ at $t = 6$ s. The abundance of the negatively charged pions, which strongly depends
on the SN temperature, reaches its maximum value 
($\sim {\cal O} (1\%)$ of the nucleon abundance) 
during the early cooling phase and at the later stage of cooling it is negligible.

Following~\cite{Lella:2022uwi}, the spectra of ALPs produced from a single core collapse SN, taking into account the gravitational redshift, is then given by
\bal
\dfrac{d N_a} {dE_a} = \int d^3 \vec{r} \int dt^* \,\alpha(r) \left(\dfrac{d^2 n^*_a}{ d E^*_a dt^*} \right)_{\rm total}.
\eal

Before ending this section, let us stress the following points: 
(i) as benchmark values for ALP-nucleon couplings, we consider $g_{an} = 0$ and $g_{ap} \ne0$,
as in KSVZ model~\cite{GrillidiCortona:2015jxo}, and (ii) we are interested in the free-streaming regime of ALP where the absorption of ALP via inverse bremsstrahlung, $a + n (p) \to  p(n) + \pi^- (\pi^+)$, and
$a+n(p) \to n(p) + \pi^0$ is negligible. Therefore
we do not consider ALP absorption while calculating the ALP flux. 

\subsection{Diffuse ALP background}
\label{sec:DSALPB}
\begin{figure}[h!]
    \centering
    \includegraphics[width=0.45\textwidth]{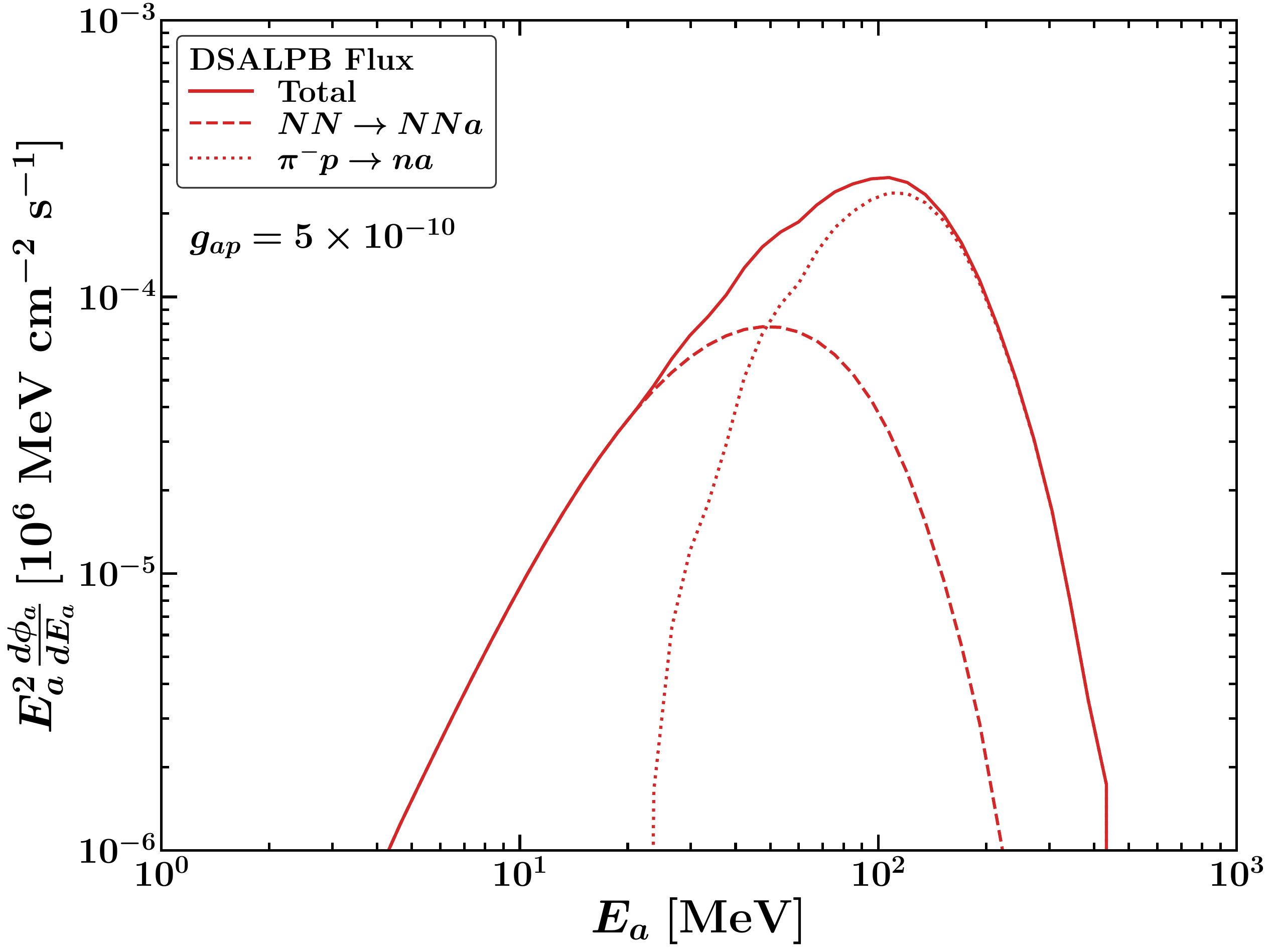}
    \caption{
    DSALPB flux arising due to the contribution of all past core collapse supernovae. We assume $\gap = 5 \times 10^{-10}$ which is below the cooling bound \cite{Lella:2023bfb}. The total DSALPB flux, considering ALP production via both \textit{NN}-bremsstrahlung and pion conversion in the PNS, is shown by the red solid line. The red dashed line and the red dotted line are the DSALPB fluxes considering only the \textit{NN}-bremsstrahlung and the pion conversion process respectively.
    }
    \label{fig:dsalpb_flux}
\end{figure}
As discussed in the previous sections, ultralight ALPs produced from the core-collapse supernovae over the entire history of our Universe give rise to a homogeneous and isotropic DSALPB ~\cite{Calore:2020tjw, Eby:2024mhd}. 
If ALPs are produced primarily by pionic processes, the DSALPB spectrum peaks at energies $E\sim {\cal O} (100\,\rm MeV)$.

According to Ref.~\cite{Calore:2020tjw}, the DSALPB flux is given by
\beq
\dfrac{d \phi_a}{d E_a}\, =\, \int_{0}^{\infty}\,(1 + z)\, \dfrac{d N_a (E_a (1 + z))}{d E_a}\,R_{\rm SN}(z)\,\left(\left|\dfrac{dt}{dz}\right|\right)\,dz,
\label{eq:dsalpb_flux}
\eeq
where $z$ is the redshift, $R_{\rm SN} (z)$ is the SN explosion
rate, taken from Ref.~\cite{Priya:2017bmm}, with a total normalization for the core-collapse rate $R_{cc} = 1.25 \times 10^{-4}\, \text{yr}^{-1}\, \text{Mpc}^{-3}$.
Furthermore, $|dt/dz|^{-1} = H_0\, (1 + z)\, [\Omega_\Lambda + \Omega_m\, (1 + z)^3]^{1/2}$ with the cosmological parameters $H_0 = 67.4\,\, \text{km s}^{-1}\text{Mpc}^{-1}$,
$\Omega_m = 0.315$, and $\Omega_\Lambda = 0.685$. 

In Fig.\,\ref{fig:dsalpb_flux}, we show the flux of diffuse supernova ALPs, $E_a^2 d \phi_a / d E_a$, from all past core collapse supernovae. For this figure, we are assuming $g_{ap} = 5 \times 10^{-10}$,
a value allowed by the SN 1987A cooling constraint, as discussed in \cite{Lella:2023bfb}.
We explicitly show the contribution of the ALP flux from $NN$-bremsstrahlung (red dashed line) and
pion conversion (red dotted line) as well as the total flux (red solid line).
We remark that, while \textit{NN}-bremsstrahlung represents the primary contribution to the DSALPB flux in the low energy regime $E_a\lesssim 100\,$MeV, the high-energy component of the spectrum is dominated by ALPs produced via pionic processes.
\begin{figure}[t!]
    \centering
    \includegraphics[width=0.48\textwidth]{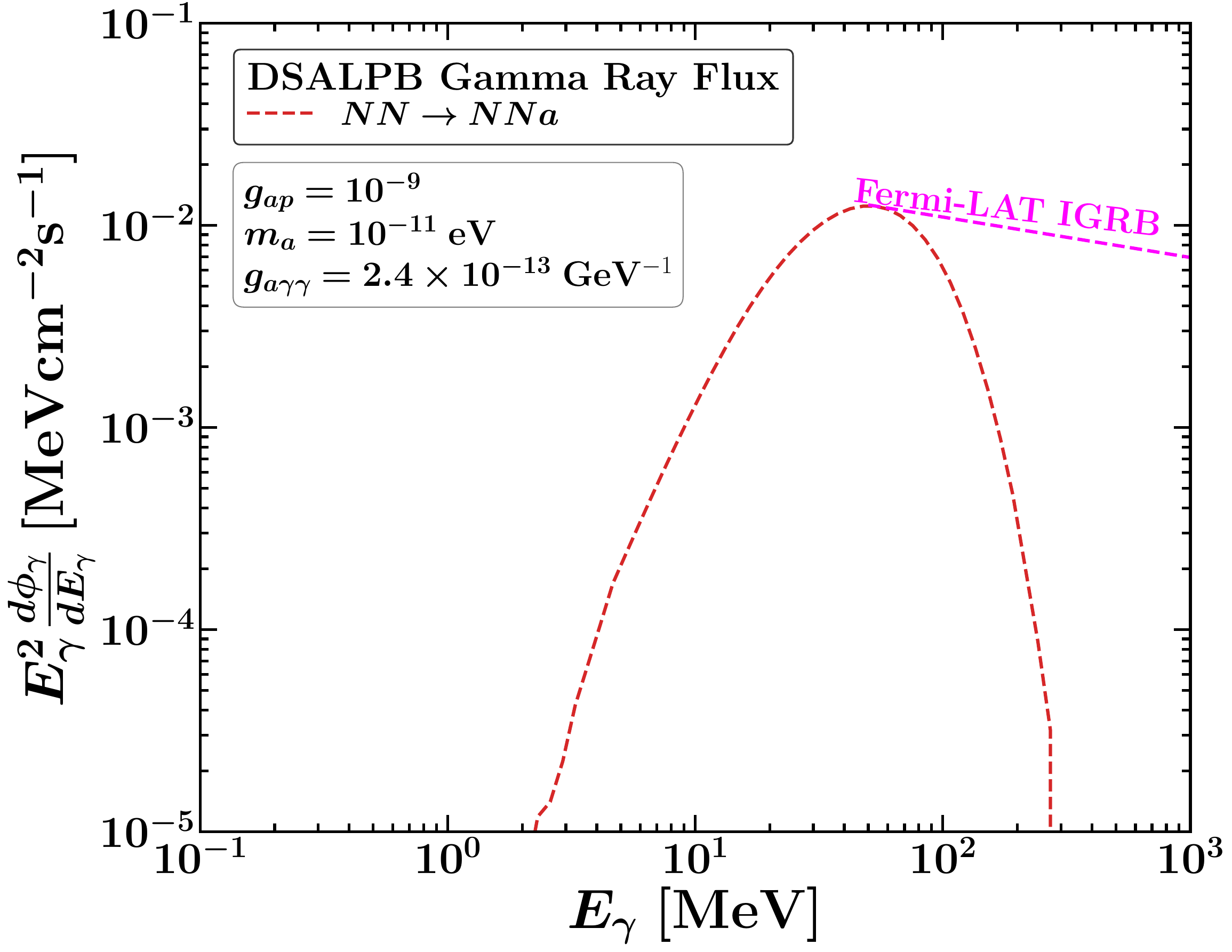}
    \includegraphics[width=0.48\textwidth]{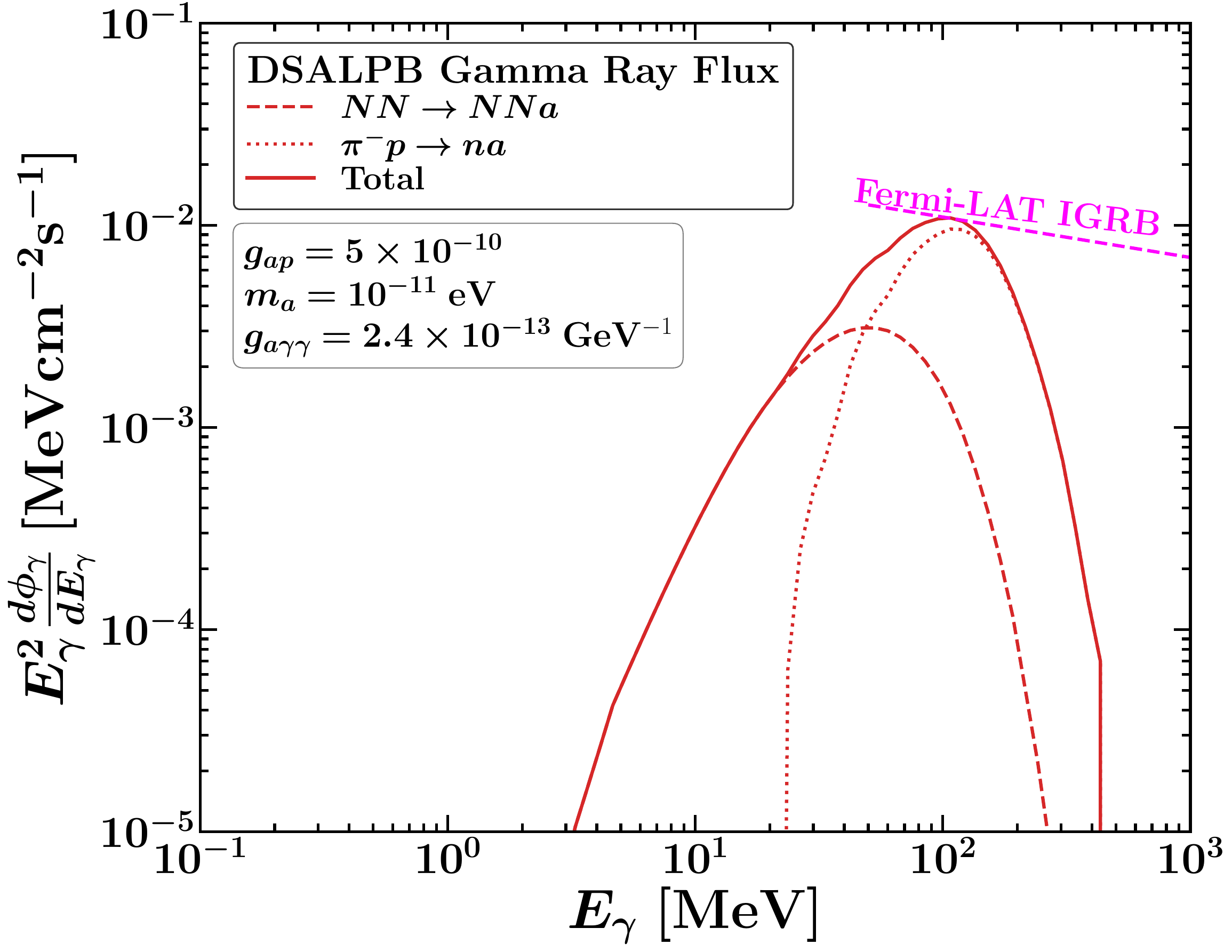}
    \caption{Diffuse photon flux as a function of photon energy. In the upper panel, we consider
    ALP production only via bremsstrahlung (red dashed line) and $g_{ap} = 10^{-9}$. In the lower panel,
    ALP production via bremsstrahlung (red dashed line), pion conversion (red dotted line)
    have been considered for $g_{ap} = 5\times 10^{-10}$ and the total flux is denoted by solid red line. In both panels, the magenta line denotes the IGRB flux measured by {\it Fermi}-LAT. 
    We fix $m_a = 10^{-11}\,\rm eV$ and $g_{a \gamma \gamma} = 2.4 \times 10^{-13}\,\rm GeV^{-1}$ in both panels.
    }
    \label{fig:diffuse_photon_flux}
\end{figure}
\section{GAMMA RAYS FROM DSALPB CONVERSION IN GALACTIC MAGNETIC FIELD }
\label{sec:gamma_ray_dsalpb}
As ALPs from the DSALPB travel toward Earth, they traverse the Milky Way’s magnetic field, where their coupling $g_{a \gamma \gamma}$ can induce oscillations into photons~\cite{Raffelt:1987im}.\footnote{Since we consider light ALPs ($m_a \lesssim 10^{-11}\,\rm eV$), the dominant photon production channel is oscillation while the decay to two photons is suppressed.} 
 
To calculate the final gamma-ray flux arising from ALP photon oscillation, we need to consider the exact structure of the galactic magnetic field and solve the ALP-photon mixing equation~\cite{Raffelt:1987im} to find the oscillation probability in the region of interest (ROI) in the sky. 
We use the \texttt{gammaALPS} package \cite{Meyer:2021pbp} to calculate the oscillation probability in the Galactic magnetic field in different
directions of the sky, $P_{a \gamma} (\ell,b)$, for given values of $E_a$, $m_a$, $g_{a \gamma \gamma}$, and take the average as
\beq
\langle P \rangle_{a \gamma} = 
\dfrac{4}{4\Delta  \Omega} \int_{0}^\pi d \ell \int_{b_{\rm min}}^{b_{\rm max}} db\,\cos b\,
P_{a \gamma} (\ell, b)\,\,,
\label{eq:osc_prob}
\eeq
where $\Delta \Omega = \int_{0}^\pi d \ell \int_{b_{\rm min}}^{b_{\rm max}} db\,\cos b$ with $\ell$ and $b$ being the galactic latitude and longitude respectively. Here we consider $b_{\rm min} = 5^{\circ}$ and $b_{\rm max} = 10^{\circ}$ in order to exclude the Galactic Center region.

Using Eq.~\eqref{eq:dsalpb_flux} and~\eqref{eq:osc_prob}, the gamma-ray flux arising due to the oscillation of DSALPB ALPs into photons while passing through the galactic magnetic field can be obtained as
\beq
\dfrac{d \phi_\gamma}{d E_\gamma}\, =\, \dfrac{d \phi_a}{d E_a}\, \langle P \rangle_{a \gamma}\:.
\label{eq:diffuse_photon_flux}
\eeq

In Fig.\,\ref{fig:diffuse_photon_flux}, we show the diffuse photon flux as a function of photon energy
and compare it with the isotropic gamma-ray background (IGRB) flux measured by {\it Fermi}-LAT which
is well-fitted by \cite{fermi,Calore:2020tjw}
\beq
\Phi(E_\gamma) =4\pi\, \times\, 2.2 \times 10^{-3} \left(\dfrac{E_\gamma}{1\,\rm MeV}\right)^{-2.2}\, \text{MeV}^{-1}\,\text{cm}^{-2}\,\text{s}^{-1}\,.
\label{eq:igrb_flux}
\eeq

In the upper panel, we consider ALP production via bremsstrahlung only and fix $g_{ap} = 10^{-9}$, corresponding to the maximum value of $g_{ap}$ allowed by SN 1987A cooling argument in the case of ALP emission via bremsstrahlung~\cite{Lella:2023bfb}. 
In the lower panel, we consider the ALP production via bremsstrahlung as well as pion conversion
for $g_{ap} = 5\times 10^{-10}$, the upper limit on $g_{ap}$ from SN 1987A cooling in the presence
of negatively charged pions in SN core \cite{Lella:2023bfb}. 
In this case, the cooling bound is stronger by a factor of 2 due to an extra ALP production channel. 
As a result, the ALP flux produced by bremsstrahlung is reduced by a factor of 4. 
In spite of this, as shown in Fig.~\ref{fig:fermi_lat_constraints}, the ALP parameter space constrained by the {\it Fermi}-LAT observations is not reduced 
since the ALPs produced from the pionic processes show a harder spectrum compared to the ALPs produced from bremsstrahlung and the {\it Fermi}-LAT effective area increases significantly at higher energies. 

\section{LIMITS AND PROJECTION ON THE
ALP-PHOTON COUPLING FROM DSALPB}
\label{sec:constraints}
\begin{figure}[t!]
    \centering
    \includegraphics[width=0.48\textwidth]{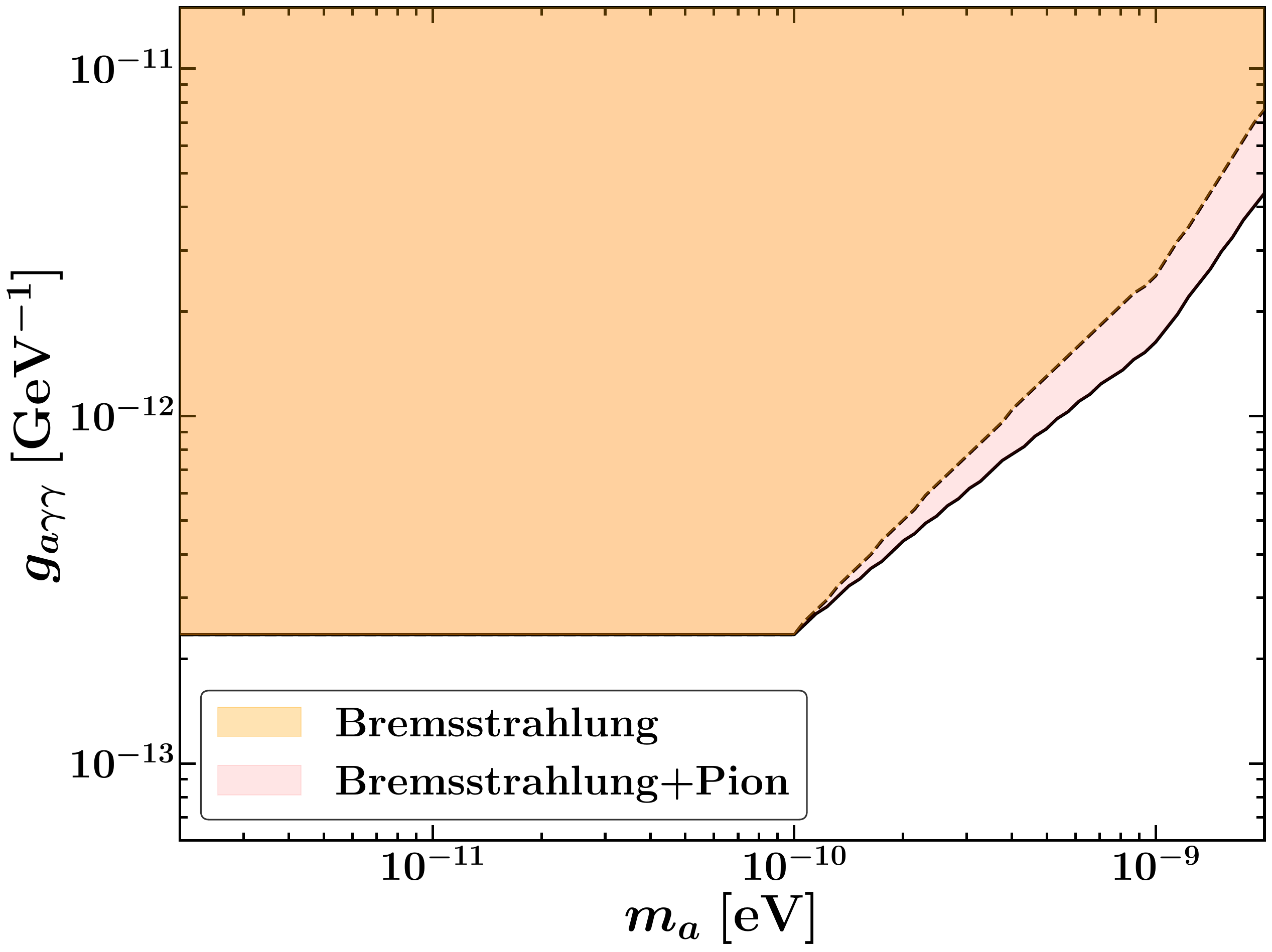}
    \caption{Constraints in $m_a - g_{a \gamma \gamma}$ plane from {\it Fermi}-LAT considering supernova ALP production via $NN$-bremsstrahlung only (orange shaded region) and $NN$-bremsstrahlung + pion conversion (pink shaded region). For deriving the constraint for the bremsstrahlung only case, we have assumed $g_{ap} = 10^{-9}$ whereas for the case considering both pion and bremsstrahlung we have assumed $g_{ap} = 5 \times 10^{-10}$. These are the values that saturate the respective SN cooling bounds.}
    \label{fig:fermi_lat_constraints}
\end{figure}

In this section, we review the ALP bound from {\it Fermi}-LAT and estimate the sensitivity of proposed future missions. Let us start from {\it Fermi}-LAT. In Fig.\,\ref{fig:fermi_lat_constraints}, we show the limit on  $g_{a \gamma \gamma}$ considering ALP production from \textit{NN}-bremsstrahlung and pion-conversion for ${g_{ap} = 5 \times 10^{-10}}$. 
For comparison, we also show the limit on  $g_{a \gamma \gamma}$
considering ALP production only from \textit{NN}-bremsstrahlung for $g_{ap} = 10^{-9}$.
For $m_a \ll 10^{-10}\,\rm eV$ both of these constraints are comparable since the ALP-photon oscillation
probability does not depend on the photon energy.
However, for $m_a > 10^{-10}\,\rm eV$, the ALP-photon oscillation probability increases with the photon energy. 
In this case, we highlight that the bound in the case of bremsstrahlung + pion processes is more constraining at higher masses than the bremsstrahlung only case. 
This is not surprising since the ALPs produced from negatively charged pions have a harder spectrum
and the ALP-photon oscillations are kept coherent up to larger masses at higher energies.

To derive the constraints from future telescopes, first
we calculate the total number of background
events within the energy range $E_{\rm min} \le E_\gamma \le E_{\rm max}$ as \cite{Lecce:2025dbz}
\bal
N_{\rm Bkg} = 4\pi \int_{E_{\rm min}}^{E_{\rm max}} \dfrac{d \Phi_{\gamma,\,\rm Bkg.}}{d E_\gamma}\,A_{\rm eff} (E_\gamma)
\,dE_\gamma\,,
\eal
where $\dfrac{d \Phi_{\gamma,\,\rm Bkg}}{d E_\gamma}$ is the background photon
flux expressed in units of $\rm cts\, MeV^{-1}\,\rm s^{-1}\,sr^{-1}$, and $A_{\rm eff}$ is the
effective area of the telescope, which is a function of the photon energy. 
We estimate $N_{\rm Bkg}$ using the data given in 
Ref.\,\cite{e-ASTROGAM:2016bph} for e-ASTROGAM, Ref.\,\cite{Caputo:2022xpx}
for AMEGO-X, Ref.\,\cite{Aramaki:2019bpi} for GRAMS-balloon and its upgraded version GRAMS-satellite,
and Ref.\,\cite{Dzhatdoev:2019kay} for MAST. The summary of the specifications of future experiments is given in Table~\ref{tab:tab1}.

Finally, we estimate the total number of signal events using the specifications of each telescope
as
\bal
N_{\rm Sig} = \int_{E_{\rm min}}^{E_{\rm max}} \dfrac{d \phi_\gamma}{d E_\gamma}
A_{\rm eff} (E_\gamma)\,dE_\gamma\,\,,
\eal
where $\dfrac{d \phi_\gamma}{d E_\gamma}$ is given in Eq.\,\eqref{eq:diffuse_photon_flux} and
we impose ${N_{\rm Sig} \ge N_{\rm Bkg}}$ to derive the upper limit
on $g_{a \gamma \gamma}$.

The projected limits from future gamma-ray telescopes in the $m_a -g_{a\gamma \gamma}$ plane are summarized in Fig.\,\ref{fig:future_constraints}. 
There, the gray shaded
region shows the combined constraints in this region of parameter space from spectral distortion of radio-quiet quasar H1821+643 by Chandra \cite{Reynes:2021bpe}, NuStar observation
from M82 \cite{Ning:2024eky}, hard x-ray observation from Betelgeuse\,\cite{Xiao:2020pra},
x-ray observation from super star clusters\,\cite{Dessert:2020lil}, and Chandra observation of
AGN NGC1275 at the center of Perseus cluster \cite{Reynolds:2019uqt}. 
 For the sake of clarity, all of these limits are derived by assuming ALP coupling only with photons. 
We find that, in the scenario in which ALPs couple also with nucleons, the constraint derived from AMEGO-X, e-ASTROGAM, GRAMS-balloon and GRAMS-satellite are not competitive with the  {\it Fermi}-LAT constraint, and that  {\it only} MAST has the potential to constrain $\gagg$ up to a few $\times 10^{-13}\,\rm GeV^{-1}$, comparable to the {\it Fermi}-LAT bound.

Finally, we also show our updated constraint from the nonobservation of ALP-induced $\gamma$-ray burst from SN 1987A.
In our scenario, we consider $g_{ap} = 5\times 10^{-10}$ and use the measurement of the $\gamma$-ray flux from SN 1987A observed by the Gamma Ray Spectrometer (GRS) on board the Solar Maximum Mission
(SMM) (see \cite{Manzari:2024jns, Lecce:2025dbz} for the limits in the scenario where $g_{a\gamma\gamma}$ and $g_{ap}$ are not independent).
Following
\cite{Brockway:1996yr,Grifols:1996id, Payez:2014xsa, Calore:2020tjw}, we consider the distance of SN 1987A $d_{\rm SN} = 50\,\rm kpc$, galactic latitude $(b)$ = $-32.1^\circ$ 
and longitude ($l$) = $279.5^\circ$ and calculate the fluence ($\cal F$) within\footnote{In our analysis, we do not consider other energy bins such as $[4.1-6.4]\,\rm MeV$ and $[10-25]\,\rm MeV$ since in these bins, ALP flux is subdominant in comparison to the flux in the energy bin
$[25,\,100]\,\rm MeV$.}
$E_\gamma = [25,\,100]\,\rm MeV$
as
\bal
{\cal F} = \dfrac{1}{4 \pi d_{\rm SN}^2}\int \left(\dfrac{d N_a}{ d E_a}\right)_{E_a = E_\gamma}
P_{a \gamma} (\ell,b) \,d E_\gamma\,\,.
\label{eq:fluence}
\eal
We impose ${\cal F} \le 0.6\,\rm cm^{-2}$ to derive the upper limit
on $g_{a \gamma \gamma}$ at 95\% confidence level, as shown
by the solid red line in Fig.\,\ref{fig:future_constraints}.

\begin{table}[h]
 \centering
\begin{tabular}{c  c  c }
\hline\\
~~Experiment~~&~~Energy range [MeV]~~&~~$N_{\rm Bkg.}$ [$\rm cts \,s^{-1}$]~~\\
\\
\hline
\\
AMEGO-X \cite{Caputo:2022xpx} & 50-200 & 1.36
\\
\\
e-ASTROGAM \cite{e-ASTROGAM:2016bph} & 50-200 & 13.38
\\
\\
MAST \cite{Dzhatdoev:2019kay} & 50-200 & 18.31
\\
\\
GRAMS-satellite \cite{Aramaki:2019bpi} & 10-100 & 1031.25
\\
\\
GRAMS-balloon \cite{Aramaki:2019bpi} & 10-100 & 390.98
\\
\\
\hline
\end{tabular}
\caption{List of future experiments and their specifications used in our analysis.}
    \label{tab:tab1}
\end{table}

Notice that the ALPs from pionic processes have energies well above the energy band $E_\gamma = [25,\,100]\,\rm MeV$ used to set the bound from SN 1987A~\cite{Payez:2014xsa}.
Thus, these ALPs do not affect the expected event counting. This is evident from the overlapping solid red and black dotted lines in Fig.~\ref{fig:future_constraints}. In the former case, ALP production from both bremsstrahlung and pion conversion is considered, whereas in the latter, only bremsstrahlung is included. Nevertheless, the presence of pions strengthens the cooling bound, forcing the ALP coupling to protons to a lower value. As a result, the overall ALP flux is reduced and the SN 1987A constraint from the Solar Maximum Mission (SMM), as shown in Fig.\,\ref{fig:future_constraints}, is slightly weaker than the bound derived in Ref.~\cite{Calore:2020tjw}.
\section{DISCUSSION AND CONCLUSIONS}
\label{sec:conclusions}
ALPs coupled to nucleons can be produced in the core of a SN via \textit{NN}-bremsstrahlung as well as through pion conversion. 
The presence of negatively charged pions makes the emitted ALP spectrum significantly harder than that produced solely through bremsstrahlung processes. 
In this work, we considered ALP production from all past core-collapse SNe, which gives rise to a diffuse ALP background with characteristic energies of ${\cal O}(100)\,\mathrm{MeV}$. 
These energetic ALPs can convert into photons in the Galactic magnetic field, generating a diffuse gamma-ray background.

After recomputing this diffuse background, including the previously overlooked contribution from the processes in SNe involving negatively charged pions, we revisited the constraints on the DSALPB using {\it Fermi}-LAT observations and present updated projections for future gamma-ray telescopes including AMEGO-X, e-ASTROGAM, GRAMS-balloon, GRAMS-satellite, and MAST. 
We find that the {\it Fermi}-LAT constraint on the ALP-photon coupling $g_{a\gamma\gamma}$ is improved for ALP masses $m_a \gtrsim 10^{-10}\,\mathrm{eV}$ when pion conversion is included as a production channel, 
as \textit{Fermi}-LAT is more sensitive to the harder spectrum predicted in this case. 
Furthermore, we show that MAST could probe values of $g_{a\gamma\gamma}$ up to  a few$\times 10^{-13}\,\rm GeV^{-1}$,
a value competitive with the {\it Fermi}-LAT constraint.

Finally, we have updated the constraint from the nonobservation of the ALP-induced gamma-ray burst associated with SN 1987A, showing that it remains approximately one order of magnitude stronger than the strongest bound we can expect from DSALPB searches. A more detailed spectral analysis over the background of the {\it Fermi}-LAT experiment, along the lines discussed in Ref.~\cite{Lella:2024hfk}, could potentially improve the constraint on the DSALPB.
Such an analysis is beyond the scope of the present work and will be addressed elsewhere.
\section{ACKNOWLEDGMENTS}
We thank Hans-Thomas Janka for giving the access to the \texttt{GARCHING} group archive. F.R.C. is supported by the Universidad de Zaragoza under the “Programa Investigo” (Programa Investigo-095-28), as part of the Plan de Recuperación, Transformación y Resiliencia, funded by the European Union-NextGenerationEU. The work of S.G. was supported by IBS under the project code IBS-R018-D1. S.G. acknowledges the hospitality of INFN-LNF
where this work was initiated.
S.G. would also like to acknowledge the hospitality of IACS, Kolkata where the final part of this work was completed.
M.G. acknowledges support from the Spanish Agencia Estatal de Investigación under Grant No. PID2019-108122GB-C31, funded by MCIN/AEI/10.13039/501100011033, and from the “European Union NextGenerationEU/PRTR” (Planes complementarios, Programa de Astrofísica y Física de Altas Energías). He also acknowledges support from Grant No. PGC2022-126078NB-C21, “Aún más allá de los modelos estándar,” funded by MCIN/AEI/10.13039/501100011033 and “ERDF A way of making Europe.” Additionally, M.G. acknowledges funding from the European Union’s Horizon 2020 research and innovation programme under the European Research Council (ERC) Grant Agreement No. ERC-2017-AdG788781 (IAXO+). T.K. acknowledges support in the form of Senior Research Fellowship (File No. 09/0080(13437)/2022-EMR-
I) from the Council of Scientific and Industrial Research
(CSIR), Government of India.
The work of A.L. was partially supported by the research grant number 2022E2J4RK ``PANTHEON: Perspectives in Astroparticle and Neutrino THEory with Old and New messengers" under the program PRIN 2022 (Mission 4, Component 1,
CUP I53D23001110006) funded by the Italian Ministero dell'Universit\`a e della Ricerca (MUR) and by the European Union – Next Generation EU.  The work
of F.M.  is supported by the European Union – Next Generation EU and
by the Italian Ministry of University and Research (MUR) via the PRIN 2022 project
No. 2022K4B58X – AxionOrigins. This article is based upon work from COST Action COSMIC WISPers CA21106, supported by COST (European Cooperation in Science and Technology).


\bibliography{refs}
\end{document}